\documentclass[twocolumn,showpacs,preprintnumbers,amsmath,amssymb,superscriptaddress]{revtex4}

\usepackage{graphicx}
\usepackage{dcolumn}
\usepackage{bm}

\begin{document}


\title{Ferromagnetism of cold fermions loaded into a decorated square lattice}

\author{Kazuto Noda}
 \affiliation{Department of Physics, Kyoto University, Kyoto 606-8502, Japan}
\author{Akihisa Koga}%
 \affiliation{Department of Physics, Kyoto University, Kyoto 606-8502, Japan}
 \affiliation{Theoretische Physik, ETH H\"onggerberg, 8093 Z\"urich, Switzerland}
 \affiliation{Department of Physics, Tokyo Institute of Technology, Tokyo 152-8551, Japan}
\author{Norio Kawakami}
 \affiliation{Department of Physics, Kyoto University, Kyoto 606-8502, Japan}
\author{Thomas Pruschke}
 \affiliation{Institut f\"ur Theoretische Physik Universit\"at G\"ottingen, G\"ottingen D-37077, Germany}
\date{\today}

\begin{abstract}
We investigate two-component ultracold fermions loaded into a decorated square lattice, which are described by the Hubbard model with repulsive interactions and nearest neighbor hoppings. By combining the real-space dynamical mean-field theory with the numerical renormalization group method, we discuss how a ferromagnetically ordered ground state in the weak coupling regime, which originates from the existence of a dispersionless band, is adiabatically connected to a Heisenberg ferrimagnetic state in the strong coupling limit. The effects of level splitting and  hopping imbalance are also addressed.
\end{abstract}

\pacs{Valid PACS appear here}
\maketitle

\section{Introduction}

Ultracold atoms have spurred intensive research activities since the successful realization of Bose-Einstein condensation \cite{Anderson,Bradley,Davis,Pethick}. Furthermore, by loading a periodic potential in the system, the optical lattice system \cite{Bloch,Bloch2} is realized, which opens a new research arena to study quantum many-body problems in condensed matter physics \cite{Geogesoptical,Lewenstein}.  This system is formed by the counterpropagating lasers, and thereby has an advantage in controlling lattice structure, dimensions, interactions, etc. Recently, great progress in reaching low temperatures for optical lattices hosting fermions has been achieved, which allows for the imaging of Fermi surfaces in three dimensions \cite{Kohl}, the observation of a Mott insulator \cite{Jordens,Schneider}, the control of superexchange interactions \cite{Trotzky,Folling}, etc. These provide an important step toward experimental observations of a magnetic phase transition in optical lattices.
Theoretically, it has already been discussed how magnetically ordered states can be realized in some bosonic \cite{Demler2002,Imambekov2003,Zhou2003,Snoek2004,Rizzi2005,Batrouni2009}
 and fermionic \cite{Dare2007,Snoek2008,Leo2008,Wang2008,Zhang2008,Higashiyama2009,Paiva2009,Gottwald2009,Mathy2009a} optical lattice systems.
(See also, continuous systems \cite{Sogo2002,Duine2005,Berdnikov2009,Conduit2009}.)

\begin{figure}[b]
\begin{center}
\includegraphics[width=7cm]{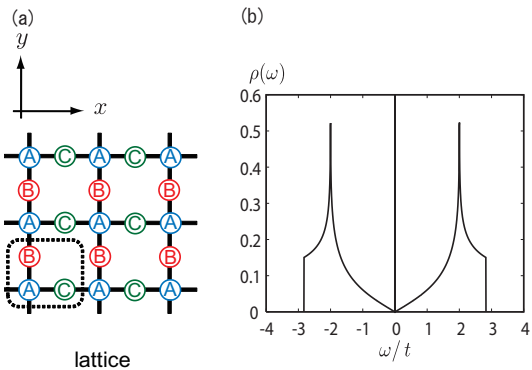}
\end{center}
\caption{
	(Color online).
	(a) Decorated square lattice. Its unit cell is shown as the dashed line.        The square lattice (A) is decorated by the B and C sublattices.
	(b) Density of states for the non-interacting case. Note that there is  a $\delta$-function peak at $\omega=0$, which reflects a dispersionless band 
 caused by a specific geometry of the lattice.}
\label{fig:DMFA}
\end{figure}

In this connection, an optical lattice with a decorated square lattice structure is particularly interesting (see Fig. \ref{fig:DMFA}(a)), where a wide variety of physics regarding magnetism and superfluidity is included. If A (B and C) sites in the system are regarded as copper (oxygen) ions, it may be reduced to a two-band model related to the high $T_c$ cuprates, where the essence of the $d$-wave superconductivity can be explored. Another interesting point related to magnetism is the ground-state property at half filling, which originates from the characteristic band structure (Fig. \ref{fig:DMFA}(b)). Due to the existence of a dispersionless band (flat band), between two dispersive bands, an infinitesimal repulsive interaction drives the system to a ferromagnetically ordered state at half filling, as guaranteed by the Lieb theorem \cite{Lieb}. 
This theorem ensures the conservation of the total magnetization in the ground state of an electron system on a bipartite lattice when the interaction is changed from infinitesimally small interaction to the strong coupling regime. This means that there is no level crossing between the ground state and low-lying eigenstates under the evolution process, which guarantees that there is no quantum phase transition at any repulsive interaction. According to this theorem, the decorated square lattice considered here has a finite magnetization at any repulsive interaction.
Therefore, this specific optical lattice may provide a unique avenue to realize the ferromagnetism in cold fermions. However, it is not clear how stable the magnetically ordered ground state is against strength of interactions, level splitting, hopping imbalance, etc, which are the accessible control parameters to drive the ferromagnetic state in an optical lattice system. This also provides us with an opportunity to study a fundamental and important issue in quantum many-body systems: what kind of crossover occurs in such a ferromagnetic state, when particle correlations are altered systematically from a weak to strong coupling regime.

To make this point clear, in this paper, we study the magnetic properties of two-component ultracold fermions on the decorated square lattice at half filling. By combining the real-space dynamical mean-field theory (R-DMFT) with the numerical renormalization group (NRG), we show how the magnetically ordered ground state is realized in the system. In particular, we elucidate that there is a smooth crossover from a flat-band ferromagnetic ground state in the weak coupling regime to a ferrimagnetic ground state in the strong coupling Heisenberg limit. 
We note here that flat band ferromagnetism is caused by a dispersionless state located just at the Fermi level, by lifting its high spin degeneracy by an infinitesimal repulsive interaction. In this sense, it is governed by the electrons only around the Fermi level (weak-coupling). On the other hand, the Heisenberg ferrimagnetism  in the strong coupling limit is stabilized by the antiferromagnetic exchange interaction induced via a virtual hopping process between the nearest neighbor sites, where electrons not only around the Fermi level but also in the high energy region contribute to the formation of magnetic moments. Therefore, when the interaction increases, the system shows a crossover from weak- to strong-coupling regimes, which is referred to as a crossover between flat-band ferromagnetism and Heisenberg ferrimagnetism in this paper. To be more specific, we can say that for infinitesimal strength of interaction, only the B and C sublattices are polarized ferromagnetically with a vanishing magnetization of the A sublattice. As the interaction increases, a magnetic moment is induced gradually on the A sublattice in an opposite direction to the ferromagnetic order (ferrimagnetism) where the total magnetization of the A, B and C sublattices is always conserved. In the following, we will explicitly show how this type of crossover behavior emerges via the calculation of static and dynamical quantities.

This paper is organized as follows. We introduce the model Hamiltonian and summarize numerical techniques in Sec. \ref{sec:Hamiltonian}. The stability of the ferromagnetically ordered state is discussed in Sec. \ref{sec:Ferromagnetism}. We also address the effects of level difference and hopping imbalance. A brief summary is given in Sec. \ref{sec:Summary}.

\section{Hamiltonian and Method}
\label{sec:Hamiltonian}

We study ultracold two-component fermions on the decorated square lattice shown in Fig. \ref{fig:DMFA}, which are described by the following Hubbard Hamiltonian,
\begin{equation}
\label{eq:Hami}
H
=
-\sum_{\langle i,j \rangle, \sigma}t_{ij}  c^{\dagger}_{i,\sigma} c_{j\sigma}
+
\sum_{i}\Delta_i n_{i}
+
U \sum_{i} n_{i\uparrow} n_{i\downarrow}
\end{equation}
where $c^{\dagger}_{i\sigma}(c_{i\sigma})$ creates (annihilates) a fermion with spin $\sigma$ at site $i$, and $n_{i\sigma} = c^{\dagger}_{i\sigma} c_{i\sigma}$. 
In our system, spin "up" and "down" just are labels to distinguish the two component fermions. Here, $t_{ij}[=t^{\rm AB} (t^{\rm AC})]$ is the nearest-neighbor hopping in the $x$ ($y$) direction, $U$ is a repulsive interaction, and $\Delta_i$ is the on-site energy at the $i$th site, which depends on the sublattice indices.

To discuss the ground state properties of the system, we use the R-DMFT, which is regarded as an extended version of dynamical mean-field theory \cite{Metzner,Pruschke,Georges}. In this method, the lattice problem is mapped onto a set of single impurity problems embedded in a self-consistent bath, where local particle correlations can be taken into account precisely. This method \cite{antiferromag} has successfully been applied to some correlated systems such as a surface \cite{Potthoff}, an interface of Mott insulator and band insulator \cite{Okamoto}, and cold fermions with a confining potential \cite{Snoek2008,Hermes,Koga}. In general, intersite correlations should be crucial to discuss the paramagnetic Mott insulating state in low-dimensional systems, which cannot be fully taken into account in the R-DMFT method. However, according to the Lieb theorem, the ground state of the Hamiltonian eq. (\ref{eq:Hami}) is always ordered except for limiting cases $U=0$ or $R=0$, where $R=t^{AC}/t^{AB}$. Therefore, it is expected that the R-DMFT, which incorporates the sublattice dependence of particle correlations, can correctly describe the ground-state magnetic properties in this system, at least, qualitatively.

In this paper, we consider a decorated square lattice system, where the unit cell is composed of three sites, as shown in Fig. 1 (a). If we exploit the naive DMFT method, in which intersite correlations are completely neglected, it is difficult to describe magnetically ordered states. However, by using the R-DMFT, which maps the original lattice to three kinds of effective impurity models in our case, we are still able to properly describe the magnetically ordered states. In particular, according to the Lieb theorem mentioned above, our system is ensured to have a finite magnetic order at any interaction strength, so that the R-DMFT provides an efficient way to investigate static and dynamical properties of the present system.
Then the Green function of the effective bath $\hat{\mathcal{G}}_{\alpha\sigma}$ is given by the Dyson equation,
\begin{eqnarray}
\label{eq:dyson}
&&
\hat{\mathcal{G}}_{\alpha\sigma}^{ -1}(\omega)
=
\nonumber
\\
&&
\left\{
	\left[
		\sum_{\bf k}
		\frac{1}{\omega + \mu - \hat{\Delta} -\hat{t}({\bf k}) 
		- \hat{\Sigma}_{\sigma}(\omega)}
	\right]^{-1}
\right\}_{\alpha\sigma}
+
\hat{\Sigma}_{\alpha\sigma}(\omega)
\nonumber
\\
\end{eqnarray}
where $\mu$ is the chemical potential, $\hat{\Sigma}_\alpha$ is the self-energy for the $\alpha$th sublattice. 
In the R-DMFT framework, we neglect the intersite correlations, which means that $\hat{\Sigma}$ is given by a diagonal matrix. The Fourier component of hopping $\hat{t}$ is given by the  following $3\times3 $ matrix, 
\begin{eqnarray}
\label{eq:hopping}
\lefteqn{\hat{t}({\bf k})=} 
\nonumber
\\
&-\left[
\begin{smallmatrix}
	0 & t^{\rm AB} (1+e^{-2i{\bf k_{y}}a}) & t^{\rm AC} (1+e^{-2i{\bf k_{x}}a}) \\
	t^{\rm AB} (1+e^{2i{\bf k_{y}}a}) & 0 & 0 \\
	t^{\rm AC} (1+e^{2i{\bf k_{x}}a}) & 0 & 0 \\
\end{smallmatrix}
\right]
\nonumber
\end{eqnarray} 
where ${\bf k}$ is the wave vector in the reciprocal lattice space and $a(=1)$ is lattice spacing. We note that the self-energy does not depend on $k$, but on the sublattice and spin indices.  This allows us to deal with intersite correlations to a certain extent through the site-dependent self-energies. The effective bath for each sublattice is explicitly given as,
\begin{eqnarray}
\label{eq:dysonA}
\mathcal{G}_{{\rm A},\sigma}^{ -1}
&=&
\left[
	\sum_{\bf k} 
	\frac{\zeta_{\rm B ,\sigma}(\omega)\zeta_{\rm C ,\sigma}(\omega)}
		 {{\rm det}({\bf k}, \omega)}
\right]^{-1}
+ \Sigma_{{\rm A},\sigma}(\omega)
\\
\label{eq:dysonB}
\mathcal{G}_{{\rm B},\sigma}^{ -1}
&=&
\left[
	\sum_{\bf k} 
	\frac{\zeta_{\rm A ,\sigma}(\omega)\zeta_{\rm C ,\sigma}(\omega) - \epsilon_{\rm AC}^2({\bf k})}
		 {{\rm det}({\bf k}, \omega)}
\right]^{-1}
+ \Sigma_{{\rm B},\sigma}(\omega)	
\nonumber
\\
\\
\label{eq:dysonC}
\mathcal{G}_{{\rm C},\sigma}^{ -1}
&=&
\left[
	\sum_{\bf k} 
	\frac{\zeta_{\rm A,\sigma}(\omega)\zeta_{\rm B,\sigma}(\omega) - \epsilon_{\rm AB}^2({\bf k})}
		 {{\rm det}({\bf k}, \omega)}
\right]^{-1}
+ \Sigma_{{\rm C},\sigma}(\omega)
\nonumber
\\
\end{eqnarray}
and
\begin{eqnarray}
{\rm det}({\bf k}, \omega)&=&
	\zeta_{\rm A,\sigma}(\omega)\zeta_{\rm B,\sigma}(\omega)\zeta_{\rm C,\sigma}(\omega)
\nonumber
\\
&&\;\;\;\;
		 - \epsilon_{\rm AC}^2({\bf k})\zeta_{\rm B,\sigma}(\omega)
		 - \epsilon_{\rm AB}^2({\bf k})\zeta_{\rm C,\sigma}(\omega)
\end{eqnarray}
where
$\epsilon_{\rm AB}({\bf k})=-2t^{\rm AB} \cos (k_y),
\:\epsilon_{\rm AC}({\bf k})=-2t^{\rm AC} \cos (k_x)$
and $\zeta_{\alpha,\sigma}(\omega)=\omega+\mu-\Delta_\alpha-\Sigma_{\alpha\sigma}(\omega)$.
By solving the three impurity problems, we obtain the corresponding self-energies. Green functions for the effective baths are then determined self-consistently through the R-DMFT equations (\ref{eq:dysonA})-(\ref{eq:dysonC}).

To discuss the ground state properties of the model system, it is necessary to solve the effective quantum impurity models in a proper way. Note here that the density of states in our system  has a singular property due to the flat band at the Fermi level, so that naive perturbation theory in $U$ may fail for the half-filled case.
Exact diagonalization and quantum Monte Carlo simulations are known to provide numerically exact results for a given system size, which are expected to reproduce some of the present results. However, these methods may not be able to resolve extremely small energy scales properly: it is not easy to correctly describe the physical properties arising from a delta-function like DOS for a flat band. In this respect, we think that Wilson's NRG (explaned below) is the most reliable method, which can treat low-energy properties with extremely high accuracy for any interaction strength, allowing us to cope with a delta-function like DOS. Since the other methods are not accurate enough to check our results at present, we confirm the validity of our results by checking them in several limiting cases.

We here use the NRG \cite{nrg_rmp} as an impurity solver. In the NRG, one discretizes the effective bath on a logarithmic mesh by introducing a discretization parameter $\Lambda(>1)$.  The resulting discrete system can be mapped to a semi-infinite chain with exponentially decreasing couplings \cite{nrg_rmp}, which allow us to discuss the properties involving exponentially small energy scales quantitatively. The resulting chain Hamiltonian is then diagonalized iteratively. An important property of this chain Hamiltonian is, that the hopping between successive sites decreases exponentially with increasing chain length. Since on the other hand the number of eigenstates increases exponentially with each iterative step, it is impossible to keep all the eigenstates computed. However, due to the decreasing coupling of the chain sites, only a restricted number of low-lying states ($=L$), usually $L=500\mathchar`-1000$ states, need to be retained at each step. The price one pays is the loss of information on high energy states at a given NRG iteration, which however can be supplied from previous iterations. As already demonstrated by a number of previous studies, the NRG is especially efficient to deal with low-energy properties with high accuracy, which is indeed the case for the present model with a flat-band structure.
To ensure that the sum rules for dynamical quantities are fulfilled, we use the complete-basis-set algorithm proposed recently \cite{Anders,Peters}. In the NRG calculations, we use a discretization parameter $\Lambda=2$ and keep $L=800$ states at each step. 
We first perform a NRG calculation for solving our impurity problems with the effective baths expressed in Eq. (3)-(5), from which we obtain the self-energy for each sublattice. The resulting self-energies are then used, in the next step, as input functions to calculate effective baths again. In the R-DMFT algorithm supplemented by the NRG, these iterative steps are continued until sufficient convergence is achieved for physical quantities.

In the following, we calculate the particle density $n_{\sigma}$ and the magnetization $m_{\alpha} = \frac{1}{2} ( n_{\alpha\uparrow}- n_{\alpha\downarrow})$ to discuss how the magnetically ordered ground state is affected by repulsive interactions, level splitting, and hopping imbalance. We also show the local density of states at each sublattice. In recent experimental developments, the density of states for gas systems becomes an observable quantity \cite{Stewart2008}. We think that this technique will be extended to fermionic optical lattice systems. This fact encourages us, therefore,  to expect that local quantities shown here will be observed in future experiments, too.
For simplicity, we set the magnetization for the A sublattice to be positive ($m_A\ge0$).

\begin{figure}[t]
\begin{center}
\includegraphics[width=7cm]{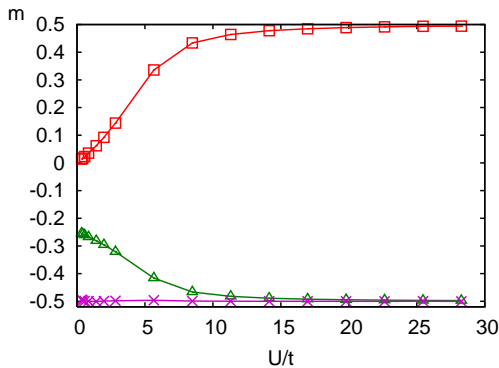}
\end{center}
\caption{
	(Color online).
Spontaneous magnetization for each sublattice: squares (triangles) correspond to the A (B and C) sublattice.  Crosses represent the total magnetization per unit cell, $m_{tot}=-0.5$, which does not depend on the interaction, in accordance with the Lieb theorem.
	}
\label{fig:correlation}
\end{figure}
\begin{figure}[htb]
\begin{center}
\includegraphics[width=8cm]{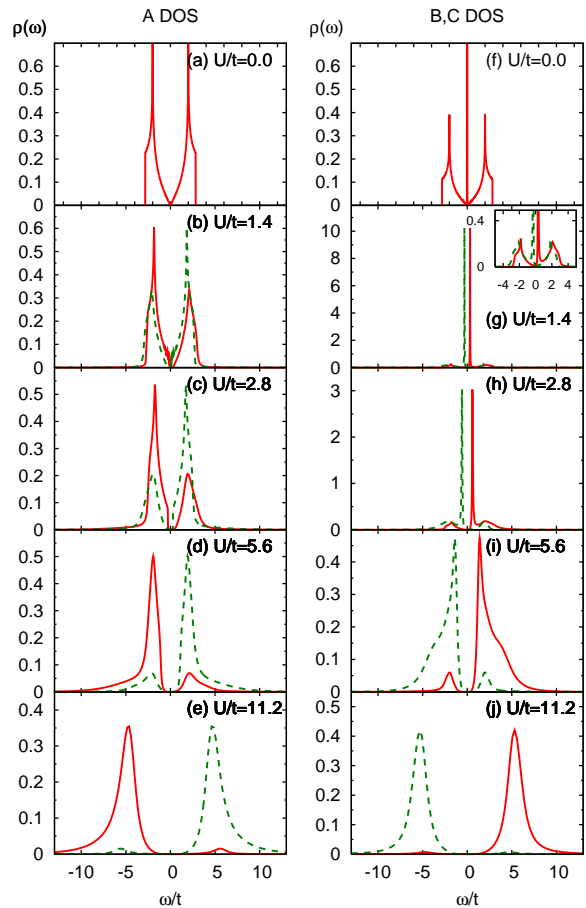}
\end{center}
\caption{
	(Color online).
Local density of states for the A sublattice (left panel) and the B, C sublattices (right panel): $U/t =$ 0.0, 1.4, 2.8, 5.6, and 11.2 (from top to bottom). Solid (dashed) lines represent the LDOS for spin-up (-down) fermions.
	}
\label{fig:correlationdos}
\end{figure}

\section{Ferromagnetism and Crossover from Weak to Strong Coupling}
\label{sec:Ferromagnetism}

First, we discuss the nature of the ferromagnetically ordered ground state for the model with isotropic hopping $t_{ij} = t$ and zero level splitting $\Delta_\alpha=0$. The spontaneous magnetization computed for each sublattice is shown in Fig. \ref{fig:correlation}. When the interaction $U$ is introduced, the magnetization for the A sublattice is induced gradually, while those for the B and C sublattices increase discontinuously from zero to $m=-0.25$ with infinitesimal interaction. This singularity is caused by the flat band in the LDOS at the Fermi level (Figs. \ref{fig:correlationdos}(a) and (f)). The resulting ferromagnetism is called {\it flat-band ferromagnetism} \cite{Lieb}. As the interaction further increases, the magnetization for each sublattice increases gradually and approaches the saturated values in the strong coupling regime, where the fermions are completely localized and the system should be described by the antiferromagnetic Heisenberg model having the nearest neighbor exchange coupling, $J \sim t^2/U$. 
In the decorated square lattice we consider here, an additional site is added at the center of each bond of the square lattice. Thus, the unit cell is composed of three sites (as shown in Fig. 1(a) ), so that in the strong coupling limit the system should be in a ferrimagnetic state. From a slightly different point of view, we can also regard our system as a 1/4 depleted square lattice, which naturally explain the emergence of a partially depleted antiferromagnetism, i.e. ferrimagnetism. Therefore, even though our lattice is bipartite, the ferrimagnetic ground state, instead of the antiferromagnetic state, is realized with staggered magnetizations ($m_A \sim 0.5$ and $m_{B,C} \sim -0.5$) in the strong coupling Heisenberg limit.
In this way, there is a smooth crossover from the flat-band ferromagnetism to the Heisenberg-type ferrimagnetism when the strength of $U$ is varied.  We have confirmed that the total magnetization of the unit cell is always preserved, $m_{tot} = -0.5$,  during the change of the interaction strength, which is consistent with the Lieb theorem.

We can clearly see in the LDOS (Fig. \ref{fig:correlationdos}) how such a crossover occurs from the weak to strong coupling regime. As mentioned above, at $U=0$ (Figs. \ref{fig:correlationdos} (a) and (f)), the LDOS has the flat band in the B, C sublattices at the Fermi level, but not in the A sublattice. When an infinitesimal interaction is introduced, the splitting of the flat band is induced at the Fermi level (Fig. \ref{fig:correlationdos} (g)), which causes the sudden increase in magnetizations for the B and C sublattices, in contrast to the smooth increase for the A sublattice (Fig. \ref{fig:correlationdos} (b)). The weight of the flat band is a third of the total weight in the DOS, resulting in the total magnetization $m_{tot} = -0.5$ per unit cell in accordance with the computed results in Fig. \ref{fig:correlation}. In this way, the ferromagnetism in the weak coupling regime is dominated by the fermions around the Fermi level in the B and C sublattices.  As $U$ increases further, the flat-band structure is gradually smeared for the B and C sublattices (Figs. \ref{fig:correlationdos} (h) and (i)), and the LDOS for A, B and C sublattices start to feature similar profiles (Figs. \ref{fig:correlationdos} (d) and (i)). Eventually, in the strong coupling regime, where the magnetization for each sublattice is almost saturated, all the fermions in the whole energy region contribute to the formation of the Heisenberg-type ferrimagnetic order.  Correspondingly, the LDOS shows similar shapes for the A, B and C sublattices (Figs. \ref{fig:correlationdos} (e) and (j)).

\begin{figure}[t]
\begin{center}
\includegraphics[width=7cm]{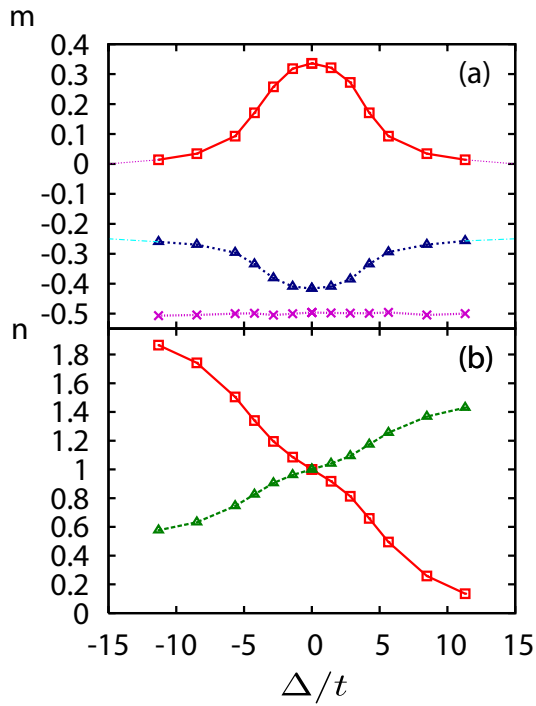}
\end{center}
\caption{
	(Color online).
(a) Spontaneous magnetization and (b) particle density of each sublattice
when $U/t = 5.6$.
Squares (triangles) represent the magnetization for the A (B or C) sublattice.
Crosses represent the total magnetization per unit cell.
	}
\label{fig:levelU2}
\end{figure}
\begin{figure}[t]
\begin{center}
\includegraphics[width=7cm]{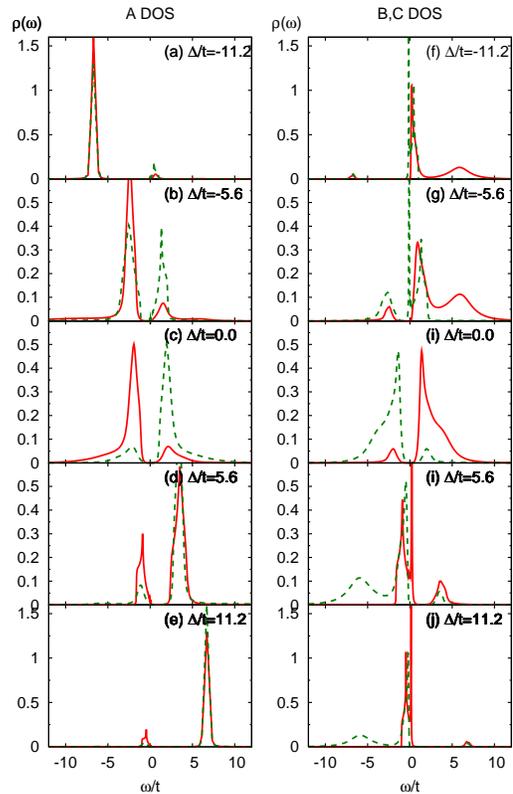}
\end{center}
\caption{
	(Color online).
Local density of states for the A sublattice (left panel) and the
B, C sublattices (right panel) when $U/t = 5.6$: $\Delta/t =$-11.2, -5.6, 0.0, 5.6, and 11.2 
(from top to bottom). Solid (dashed) lines represent the LDOS for 
up (down) spin fermions.  At $\Delta/t =$ 11.2 (-11.2), a $\delta$-function like peak appears for up-spin (down-spin) fermions in the B,C sublattices. 
	}
\label{fig:levelU2dos}
\end{figure}

We next consider the effect of the level splitting between the A and (B, C) sublattices, $\Delta (=\Delta_A-\Delta_{B, C})$, in the half-filled system. Such a splitting may be manipulated experimentally by introducing a site-dependent potential. By performing the R-DMFT calculations with the total number of particles fixed, we obtain the results for a specific choice of the interaction, $U/t=5.6$, which are shown in Fig. \ref{fig:levelU2}. Note that the obtained magnetizations  are symmetric about $\Delta=0$. When $\Delta=0$, the system is in the region of ferrimagnetic ground state, where the sublattice magnetization is large enough $(|m_A|\sim 0.3$ and $|m_{B,C}\sim 0.4|)$. The introduction of the level splitting $\Delta$ decreases (increases) the particle density for the sublattice A (B and C), as shown in Fig. \ref{fig:levelU2dos}. At the same time, the magnetization for each sublattice is monotonically decreased, and the effect of onsite correlations becomes less important. When $|\Delta|/t=11.2$, the sublattice magnetization for the A sublattice almost vanishes, but is finite for the B and C sublattices. This implies that the crossover occurs from the ferrimagnetic state to the flat-band ferromagnetic state when $|\Delta|$ is increased. The crossover is indeed confirmed by observing LDOS calculated for the same parameters shown in Fig. \ref{fig:levelU2dos}. It is seen that a $\delta$-function like peak characteristic of the flat-band ferromagnetism appears in the vicinity of the Fermi level only in the LDOS for the B and C sublattices at $|\Delta|/t\sim 11.2$.  Namely, the magnetization is governed by fermion states near the Fermi level for the B and C sublattices. 

We would like to comment here on a technical problem in our numerical calculations, which arises when we numerically fix the total number of particles in the presence of large level difference $\Delta$. When the level difference is introduced, the DOS does not preserve particle-hole symmetry, which was used efficiently to fix the particle number at half-filling for $\Delta= 0$. Therefore, the introduction of finite $\Delta$ requires more elaborate numerical calculations to keep the filling factor unchanged with sufficient accuracy. Thus DMFT calculations typically take much longer time to adjust the chemical potential properly in each iteration step. For larger level splitting, the ground state is very sensitive to the change in the chemical potential, so that this sort of problem becomes more serious. Nevertheless, the half-filled system we address here satisfies the condition of the Lieb theorem for any value of $\Delta$, which ensures that the magnetic state should be stable even for very large $|\Delta|$, although the ground state may become very sensitive to a small amount of particle or hole doping. The results deduced numerically for large $|\Delta|$ are indeed consistent with the arguments based on the Lieb theorem.

\begin{figure}[t]
\begin{center}
\includegraphics[width=7cm]{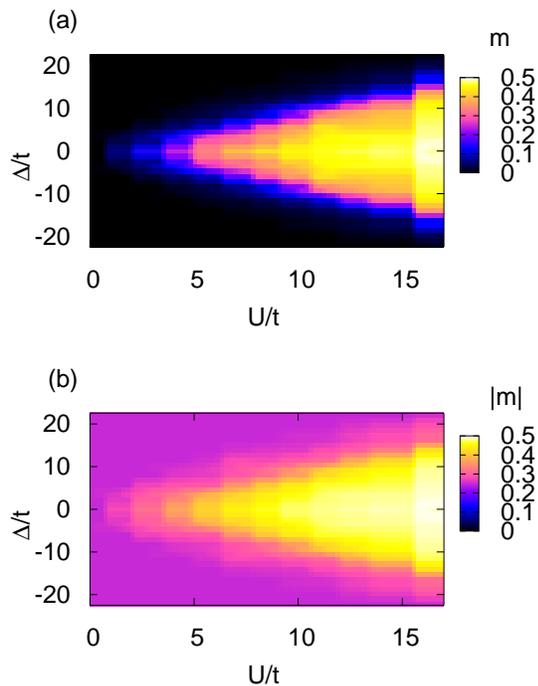}
\end{center}
\caption{
	(Color online).
 Spontaneous magnetization for (a) A sublattice (b) B and C sublattices 
as functions of $U/t$ and $\Delta/t$. Bright (dark) region has a large (small) magnetization. It is seen that a crossover appears around the lines of  $U \sim |\Delta|$ and $U \sim W$.
	}
\label{fig:level}
\end{figure}
To discuss the crossover between two extreme limits of the magnetically ordered state, we systematically perform similar calculations for various choices of the level splitting and the interaction. The obtained results for the magnetization are shown (shading plots) in Fig. \ref{fig:level}. When $U/t$ is large and $|\Delta|/t$ is small, the onsite Coulomb interaction plays an important role in stabilizing the ferrimagnetically ordered ground state with $|m_\alpha| \sim 0.5$, as mentioned above. On the other hand, when the level splitting is large, the magnetization for the A sublattice vanishes and those for the B and C sublattices are close to half of full polarization, $m_{B, C}\sim - 0.25$, characteristic of flat-band ferromagnetism. It is seen  in Fig. \ref{fig:level} that the crossover between two types of magnetically ordered states appears around $|\Delta/U|\simeq1$ and $U/W\simeq1$, 
where $W$($=4\sqrt{2}t$) means the bandwidth at $\Delta=0$.
Let us comment on the crossover as a function of the level difference $\Delta$ found here. This crossover is related to the original one as a function of $U$ 
discussed above. In fact, the physics emerging in both cases are essentially the same. The crucial point is that a flat band always exists for any value of the level difference $\Delta$ at $U=0$, and it consists of a specific combination of the states belonging to the B and C sublattices. For finite $\Delta$, the strength of the effective interaction should be compared with the value of $\Delta$: for a given value of $U$, larger (smaller) $\Delta$ effectively drives the system to a weak (strong) coupling regime. Therefore the change in the level difference causes the change in the effective interaction, which in turn gives rise to a crossover from weak- to strong-coupling regimes. One can indeed see the formation of a delta-function like peak in Fig. 5(i) and (j) for large $\Delta$, which is characteristic of the weak-coupling regime (flat-band ferromagnetism). We note that the Lieb theorem even in this case guarantees no transition under the change of $\Delta$

\begin{figure}[t]
\begin{center}
\includegraphics[width=7cm]{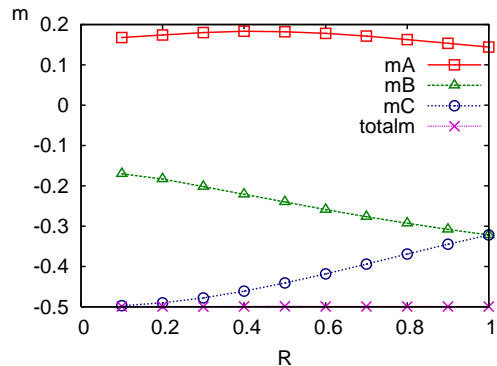}
\end{center}
\caption{
	(Color online).
	Spontaneous magnetizations as functions of $R(=t^{\rm AC}/t^{\rm AB})$ 
	when $ U/t^{\rm AB} = 2.8$.
	}
\label{fig:imbalance}
\end{figure}

We finally discuss the effects of hopping imbalance $R(=t^{\rm AC}/t^{\rm AB})$, which may be tunable by laser techniques. When $R$ is decreased from unity, hopping of fermions becomes spatially anisotropic. For example, in the small $R$ region, the system can be regarded as the one-dimensional Hubbard chains in $y$-direction that are weakly coupled to isolated sites (C sublattice). In the case, the absolute values of the magnetizations for the A and B sublattices seem to approach the same finite value ($|m_{A, B} |\sim 0.15$) as $R$ decreases, as shown in Fig. \ref{fig:imbalance}. On the other hand, the spins for the C sublattice become fully polarized, since those spins become almost free. Note that  as far as $R\neq 0$, the system satisfies the condition of the Lieb theorem, and thereby the total magnetization, $m_{tot} = -0.5$, should not be altered during the change of $R$, as is indeed the case in  Fig. \ref{fig:imbalance}. In the special case of $R=0$, the system is completely decoupled into two subsystems consisting of Hubbard chains and isolated sites. 
The Hubbard chain realized at $R=0$, which consists of A and B sublattices, corresponds to a 1D system with periodic boundary conditions. However, in the DMFT the information of the system is incorporated through the shape of the density of states. So, it is rather difficult to deduce physical properties correctly in this limiting case. This point thus remains an open issue for future studies. In the Hubbard chain, a magnetic long range order does not appear even at absolute zero, and the associated elementary spin excitations are gapless. Therefore, it might be expected that on the introduction of infinitesimal $R$, isolated free spins (C sublattice) affect the Hubbard chains to induce finite sublattice magnetizations in the A and B sublattices. Since our mean-field calculations may not be appropriate to discuss the weakly coupled chains, magnetic properties around $R\sim0$ should be examined by incorporating intersite correlation more precisely, which will be addressed elsewhere.

Before closing this section, we would like to comment on the validity of our study. In the framework of the R-DMFT, we have fully incorporated onsite correlations, but have not dealt with intersite correlations sufficiently. This indeed leads to the sublattice magnetizations that are somewhat overestimated, particularly in the strong coupling regime: the magnetization for each sublattice takes almost saturated value, $|m_{A,B,C}|\sim 0.5$, which should be suppressed if we introduce quantum fluctuations originating from intersite correlations more precisely. 
However, we have confirmed that our calculations always satisfy the condition 
of the Lieb theorem that the total magnetization is constant, $m_{tot} = -0.5 $, in the whole parameter space. Note that the R-DMFT calculation of the ground state is performed independently for various choices of interactions, from which we figure out how large the ground-state magnetization is. The above confirmation is very important because the R-DMFT is a mean field theory and therefore it is not a priori obvious whether this method is applicable to the flat-band system considered here. We have also revealed that the crossover in the density of state interpolates correct behaviors both at weak and strong coupling limits. This result also encourages us to believe the validity of the method. 
Therefore, we believe that our R-DMFT study can describe the essential properties of the system at least qualitatively in the whole parameter region, although intersite magnetic correlations should be taken into account in order to discuss the results more quantitatively in the strong coupling regime and for $R\to0$.

\section{Summary and Discussions}
\label{sec:Summary}
We have investigated the nature of the ferromagnetically ordered ground state 
in the half-filled Hubbard model on a decorated square lattice by means of R-DMFT and NRG. By varying the interaction strength, the level splitting, and the hopping imbalance, we have found that the ferromagnetic state, which originates from a flat-band structure, is adiabatically connected to the Heisenberg ferrimagnetic state in the strong coupling regime. 
The magnetic ground state becomes more stable as $U$ increases from zero, but beyond the crossover regime it is governed by the Heisenberg exchange interaction, $ \sim t^2/U$, whose magnitude decreases as $U$ further increases. Therefore, the energy gain forming the magnetically ordered state is largest in the crossover regime, which implies that the magnetic ground state should be most stable in the crossover regime.
To realize the above magnetic state experimentally in optical lattice systems, it is important to examine how stable it is at finite temperatures.  In the pure 2D case discussed here, a finite-temperature phase transition does not occur according to the Mermin-Wagner theorem. 
If we consider a quasi-2D system (e.g. a collection of weakly coupled layers, etc), however, a phase transition can occur at a finite temperature.
The corresponding transition temperature is expected to be highest in the crossover regime, according to the above discussions on the stability of the ground state. This fact encourages experimental investigations on a quasi-2D optical lattice system, for which the system parameters are tuned in the crossover regime. 

Another point to be noticed in optical lattice systems is that there is a confining trap for ultracold atoms, which superposes a smoothly varying harmonic potential to the periodic potential for the decorated square lattice. In a local density approximation, it can be regarded as a spatially modulated chemical potential. Therefore, for small $U$, the area which satisfies the half-filling condition relevant to forming a local Mott state, if it exists,  may be very small around the center of the system. As $U$ increases, the area of local Mott state is extended, and at the same time, the ferromagnetic state becomes more stable, changing its character from a flat-band type to a Heisenberg type gradually. Therefore, it is expected that for reasonably large interactions around the crossover regime, the ferromagnetically ordered state may be observed in the optical lattice. Fortunately, this conclusion is consistent with the temperature effects mentioned above. Detailed discussions on  a trap potential will be addressed elsewhere.

\begin{acknowledgments}
The authors would like to thank Y. Koyama  for
valuable discussions.
This work was partly supported by the Grant-in-Aid for Scientific Research 
[20029013, 21540359  (N.K.) and 20740194 (A.K.)] and 
 the Grant-in-Aid for the Global COE Programs 
"The Next Generation of Physics, Spun from Universality and Emergence" 
and "Nanoscience and Quantum Physics"
from the Ministry of Education, Culture, Sports, 
Science and Technology (MEXT) of Japan.
TP would like to acknowledge the hospitality and 
financial support of the Department of Physics of Kyoto University.
\end{acknowledgments}

\end{document}